\documentclass[12pt,preprint]{aastex}

\shorttitle{Ultraviolet-Luminous Galaxies}
\shortauthors{Heckman et al.}

\begin{document}


\title{The Properties of Ultraviolet-Luminous Galaxies at the Current Epoch}


\author{Timothy M. Heckman\altaffilmark{1},
Charles G. Hoopes\altaffilmark{1},
Mark Seibert\altaffilmark{2},
Christopher Martin\altaffilmark{2},
Samir Salim\altaffilmark{3},
R. Michael Rich\altaffilmark{3},
Guinevere Kauffmann\altaffilmark{4},
Stephane Charlot\altaffilmark{4,5},
Tom A. Barlow\altaffilmark{2},
Luciana Bianchi\altaffilmark{1}, 
Yong-Ik Byun\altaffilmark{6}, 
Jose Donas\altaffilmark{7},
Karl Forster\altaffilmark{2}, 
Peter G. Friedman\altaffilmark{2},
Patrick N. Jelinsky\altaffilmark{8},
Young-Wook  Lee\altaffilmark{6},
Barry F. Madore\altaffilmark{9,10}, 
Roger F. Malina\altaffilmark{7}, 
Bruno Milliard\altaffilmark{7},
Patrick F. Morrissey\altaffilmark{2}, 
Susan G. Neff\altaffilmark{11},
David Schiminovich\altaffilmark{2},
Oswald H. W. Siegmund\altaffilmark{8}, 
Todd Small\altaffilmark{2},
Alex S. Szalay\altaffilmark{1}, 
Barry Y. Welsh\altaffilmark{8},
and Ted K. Wyder\altaffilmark{2}}

\altaffiltext{1}{Center for Astrophysical Sciences, Department of Physics \&
Astronomy, 
Johns Hopkins University, 3400 N. Charles St., Baltimore, MD, 21218}

\altaffiltext{2}{Division of Physics, Mathematics, and Astronomy, California Institute of Technology, Pasadena, CA, 91125}

\altaffiltext{3}{Department of Physics and Astronomy, University of California 
at Los Angeles, 8965 Mathematical Sciences Building, Los Angeles, CA 90095}

\altaffiltext{4}{Max Planck Institute for Astrophysics, Garching, Germany}

\altaffiltext{5}{Institut d'Astrophysique de Paris, 98 bis boulevard Arago, 750
14 Paris, France}

\altaffiltext{6}{Center for Space Astrophysics, Yonsei University, Seoul
120-749, Korea}

\altaffiltext{7}{Laboratoire d'Astrophysique de Marseille, BP 8, Traverse
du Siphon, 13376 Marseille Cedex 12, France}

\altaffiltext{8}{Space Sciences Laboratory, University of California at
Berkeley, 601 Campbell Hall, Berkeley, CA 94720}

\altaffiltext{9}{Observatories of the Carnegie Institution of Washington,
813 Santa Barbara St., Pasadena, CA 91101}

\altaffiltext{10}{NASA/IPAC Extragalactic Database, California Institute
of Technology, MC 100-22, 770 S. Wilson Ave., Pasadena, CA 91125}

\altaffiltext{11}{Laboratory for Astronomy and Solar Physics, NASA Goddard
Space Flight Center, Greenbelt, MD 20771}



\begin{abstract}
We have used the first matched set of GALEX and Sloan Digital Sky Survey (SDSS)
data to investigate the properties of a sample of 
74 nearby ($z < $ 0.3)
galaxies with far-ultraviolet luminosities greater than
$2 \times 10^{10} L_{\odot}$, chosen to overlap the luminosity range
of typical high-z Lyman Break Galaxies (LBGs).
GALEX deep surveys have shown that ultraviolet-luminous galaxies
(UVLGs) similar to these are the fastest evolving 
component of the UV galaxy population. 
Model fits to the combined 
GALEX and SDSS photometry yield typical FUV extinctions in UVLGs of
0.5 to 2 magnitudes
(similar to LBGs and to less luminous GALEX-selected galaxies).
The implied star formation rates are SFR $\sim$ 3 to 30 $M_{\odot}$/year.
This overlaps the range of SFRs for
LBGs. We find a strong inverse correlation between galaxy mass and 
far-ultraviolet
surface brightness, and on this basis divide the sample into 
``large'' and ``compact'' UVLGs.
The large UVLGs are
relatively massive ($M_* \sim 10^{11} M_{\odot}$) late type disk galaxies
forming stars at a rate similar to their past average 
($M_*/SFR \sim t_{Hubble}$). They are metal rich ($\sim$ 
solar), have
intermediate optical-UV colors ($FUV-r \sim$ 2 to 3), and about a third
host a Type 2 (obscured) Active Galactic Nucleus. In contrast, the compact
UVLGs have half-light radii of a 
few kpc or less
(similar to LBGs ). They are relatively low-mass galaxies 
($M_* \sim 10^{10} M_{\odot}$) with typical velocity dispersions of
60 to 150 km/s.
They span a range in metallicity from $\sim$ 0.3 to 1 times solar, have
blue optical-UV colors ($FUV-r \sim$ 0.5 to 2), and are forming stars at a rate 
sufficient to build the present galaxy in $\sim$1 - 2 Gigayear. In all these
respects they appear similar to the LBG population. These 
``living fossils'' may therefore provide an opportunity for detailed 
investigation of the physical processes occurring in typical star forming
galaxies in the early universe.
\end{abstract}




\section{Introduction}

Over the past several years the first picture of the cosmic evolution
of the global star-formation rate has been sketched (e.g. Giavalisco et
al. 2004; Dickinson et al. 2003).
This is a truly
remarkable achievement.
However, there is considerable work to be done.  The ultimate goal of
course is to understand the astrophysical processes that drove the
evolution of star formation and to relate these processes to the
building of galaxies, the fueling of active galactic nuclei, and the
development of large-scale structure.  This will require detailed and
statistically robust comparisons of the fundamental properties of star
forming galaxies as a function of redshift. 

One obstacle has been that
our understanding of the evolution of star formation has until
recently been a patchwork quilt that relied on different techniques
(each with their own selection biases and systematic methodological
uncertainties) over different redshift ranges. For example, our most
detailed information about star formation in the early universe has
been provided by the large sample of (rest-frame) UV-selected Lyman
Break Galaxies (LBGs). Ironically, the properties of the UV-selected
galaxy population have been much more fully characterized at z $\sim$
3 than at $z \sim$ 0 (e.g. Shapley et al. 2001; Papovich et al. 2001;
Giavalisco 2002 - hereafter S01, P01, and G02 respectively).

This situation has changed dramatically with the successful launch of
the {\it Galaxy Evolution Explorer} (GALEX). Its All-sky Imaging
Survey (AIS) and Medium Imaging Survey (MIS) will generate UV-selected
samples of a million relatively nearby star forming galaxies (Martin
et al 2004) with optical photometry and
spectroscopy from the Sloan Digital Sky Survey (SDSS - York et
al. 2000).

One of the first major results from the GALEX mission is the
measurement of the evolution of the galaxy ultraviolet luminosity
function from $z \sim$ 0 to 4 (Arnouts et al. 2004; Schiminovich et
al., 2004).  As emphasized in these papers, the rate of cosmic
evolution in the ultraviolet galaxy population is strongest at the
highest luminosities. In particular, the co-moving space density of
galaxies with far-ultraviolet luminosities $L_{FUV}> 10^{10.1}
L_{\odot}$ increases by a factor of $\sim$30 from $z$ = 0 to 1, and by
additional factor of $\sim$4 out to $z$ = 3.

Thus, it is clearly important to identify and study the most
ultraviolet- luminous galaxies in the relatively nearby universe, and
then compare the properties of these (rare) objects to their
counterparts at high-redshift (most notably, the LBGs). In the present
paper we use the results from the initial matching of the GALEX and
SDSS data to initiate such an investigation.

\section{Sample Selection and Data Description}

Our goal is to define a sample of local galaxies that overlap significantly
in UV luminosity with LBGs. However, within this range,
the number of local galaxies is falling very steeply with luminosity.
For this reason, we have chosen to define a sample of
``Ultraviolet Luminous Galaxies'' (UVLGs) as having
FUV luminosities ( $\lambda
P_{\lambda}$ at 1530 \AA) $\geq 2 \times 10^{10}
L_{\odot}$.\footnote {We use $H_0$=70 km sec$^{-1}$ Mpc$^{-1}$,
$\Omega_m$=0.3, and $\Omega_{\Lambda}$=0.7 throughout this paper.}
This defining luminosity is roughly half way between the characteristic ($L_*$)
value for the present day UV galaxy luminosity function
($10^{9.6} L_{\odot}$ - Wyder et al. 2004) and that of the LBG
population at $z \sim$ 3 ($10^{10.8} L_{\odot}$ -
G02; Arnouts et al. 2004). 

We started by matching the GALEX IR0.2 and SDSS Data Release One catalogs (as
detailed in Seibert et al.  2004). We then selected those objects
spectroscopically classified as galaxes by the SDSS, specifically
excluding objects spectroscopically classified QSOs (or Type 1 Seyfert
galaxies).
Finally, we have inspected the spectra and removed a few weak Type 1
AGN that had been misclassified by SDSS. This gives us a sample of 74
galaxies, spread rather uniformly in redshift between 0.1 and 0.3
(median $z$ = 0.19). These galaxies are indeed rare:
the co-moving space density of UVLGs 
at the current epoch is only $\sim10^{-5}$ Mpc$^{-3}$, 
several hundred times smaller than the co-moving density of LBG at $z\sim$3.

The SDSS data by themselves provide a host of important galaxy
parameters (Abazajian et al. 2003; 
Stoughton et al. 2002).  We have used the SDSS Data Release Two
archive to retrieve the Petrosian magnitudes and seeing-deconvolved
half-light radii in the
$u$ and $r$ bands, the concentration index $C$ in the r-band, and the
redshift for all the UVLGs. For the UVLGs in the SDSS "main" galaxy
sample (Strauss et al. 2002), we
have used the value-added catalogs described
at http://www.mpa-garching.mpg.de/SDSS/ to retrieve estimates of
gas-phase metallicities, emission-line widths,
age-sensitive 4000 \AA\ break amplitudes, 
emission-line ratios, and classifications (AGN vs.  star forming). See
Kauffmann et al (2003a,b), Brinchmann et al. (2004), and Tremonti et
al. (2004) for details. For the other UVLGs we have used the
methodology described in Tremonti et al. (2004) to measure fluxes
and widths of the emission-lines.
The combined GALEX plus SDSS 7-band photometry has been used in
conjunction with an extensive grid of models to estimate the stellar
masses, FUV extinctions, and star formation rates.
See Salim
et al. (2004) for details.
\footnote{A table with the relevant SDSS and GALEX-derived parameters for the 
74 sample galaxies is available
at http://www.mpa-garching.mpg.de/SDSS/Data/uvlg.html}

\section{Results}

\subsection{Basic Structural Properties}

In Figure 1 we show a plot of the far-UV luminosity $L_{fuv}$ {\it
vs.}  the half-light radius in the SDSS $u$-band ($R_u$). 
\footnote{We have used radii based on the best fitting seeing-convolved
exponential disk models from the SDSS
(see http://www.sdss.org/dr2/algorithms/photometry.html\#mag\_model).
While these radii are therefore
corrected in principle for the effects of seeing,
it is important to note that the compact
UVLGs are not well-resolved in the SDSS images.
Thus, the radii and implied
surface brightnesses of the compact UVLGs are relatively uncertain.}
For the redshift range of the UVLGs, the SDSS $u$-band covers
the rest-frame range $\sim$2700 to 3200\AA\ band, where the contribution
by young stars should dominate (given the
high specific star formation rates discussed below).
The UVLGs
span a broad range in size, from relatively compact systems with radii
of a kpc (similar to LBGs - Ferguson et al. 2004) to very large
galaxies with radii of over 10 kpc. This implies a correspondingly large
range in far-UV surface brightness ($I_{fuv}$ - Figure 2). The UVLGs overlap
the range of $I_{fuv}$ of typical LBGs ($\sim 10^9$ to $10^{10}
L_{\odot}$ kpc$^{-2}$), but most are significantly fainter. Figure
2 shows that there is a strong inverse correlation between surface
brightness and mass.

While it is clear that a continuum of properties is present,
the wide range of physical properties makes it instructive to subdivide
our sample into two
categories: ``large'' ($I_{fuv} < 10^8 L_{\odot}$ kpc$^{-2}$) and
``compact'' ($I_{fuv} > 10^8 L_{\odot}$ kpc$^{-2}$). This choice
not only divides the sample into two subsets with roughly equal
numbers, but it also divides the sample in galaxy mass at the
critical value ($M_* \sim 10^{10.5} M_{\odot}$) where
Kauffmann et al. (2003c) find that the SDSS galaxy population
abruptly transitions from {\it primarily}
low mass, low density, low concentration,
disk dominated galaxies with young stellar populations to
high mass, high density, high concentration, bulge dominated systems
with old stellar populations.

The compact, high surface-brightness UVLGs are typically low-mass
galaxies ($M_{*} \sim 10^{9.5}$ to $10^{10.7} M_{\odot}$). This mass
range is very similar to that of LBGs (S01,P01,G02).
These compact
UVLGs are larger, more luminous, and more massive than local HII
galaxies (Telles \& Terlevich 1997) and overlap the high mass range of
compact galaxies found at higher redshift (0.4 $\leq z \leq$ 1) in the
Hubble Deep Field (Phillips et. al 1997). 

The large, low surface-brightness UVLGs are significantly more massive
($M_{*} \sim 10^{10.5}$ to $10^{11.3} M_{\odot}$).
The high-mass galaxies in the SDSS that do have
a significant population of young stars have the low concentrations
typical of disk galaxies ($C <$ 2.6), and this is the case for the
large UVLGs (90 \% of which have $C <$ 2.6). 
The large UVLGs have typical values of 
effective stellar surface mass densities $\mu_* \sim
10^8$ to $10^{8.5} M_{\odot}$ kpc$^{-2}$. This is about a factor
of three lower than average for comparably massive late type galaxies
(see Figure 14 in Kauffmann et al 2003c). These UVLGs are corresponding
larger than average late type galaxies of the same mass (Shen et al. 2003).
These results are not surprising, since Kauffmann et al. (2003c)
showed that at fixed galaxy mass, SDSS galaxies with lower $\mu_*$
(and thus larger radii) have younger
stellar populations.

\subsection{Star Formation Rates \& Related Parameters}

The typical values for the FUV extinction in UVLGs are
modest: $A_{FUV} \sim$ 0.5 to 2 magnitudes. This range is
similar to that of a NUV-selected sample of
galaxies that is representative of the local GALEX population
(Buat et al. 2004), as well as that of LBGs (S01,P01). Adopting a
Kroupa (2001) Initial Mass Function, the typical estimated star
formation rates for the UVLGs are $\sim$ 3 to 30 $M_{\odot}$
year$^{-1}$ (or $\sim$ 5 to 50 $M_{\odot}$ year$^{-1}$ adopting the
historically standard Salpeter Initial Mass Function). These star
formation rates overlap the range spanned by LBGs (S01, P01, G02).

In Figure 3 we show that the {\it specific} star formation rate
($SFR/M_*$) is a strong function of the far-UV surface brightness.
The compact, high surface brightness UVLGs have typical values of
$SFR/M_* \sim 10^{-8.6}$ to $10^{-9.8}$ year$^{-1}$, compared to $\sim
10^{-9.5}$ to $10^{-10.5}$ year$^{-1}$ for the large low surface
brightness UVLGs. Of course the specific star formation rate is
roughly the inverse of the time it would take to form the present
galaxy mass at the present star formation rate. Thus, the large UVLGs
are (in the mean) forming stars at a rate similar to their past rate
averaged over a Hubble time. In contrast, the compact UVLGs have a
typical ``galaxy-building time'' of only 1 to 2 Gigayear. Thus, they
are ``starburst'' systems. The {\it most}
compact UVLGs ($I_{fuv} \geq 10^9 L_{\odot}$ kpc$^{-2}$)
have galaxy-building times of only 0.1 to 1 Gyr, very similar to the
corresponding timescales in LBGs (S01,P01,G02).

Salim et al. (2004) and Brinchmann et al. (2004) have shown that the
specific star formation rates in galaxies correlate strongly with the
FUV-optical colors and the 4000 \AA\ break amplitude (D4000),
respectively. This is true for the UVLGs. The large UVLGs have
intermediate colors ($FUV-r \sim$ 2 to 3) and 4000 \AA\ break
strengths (D4000 $\sim$ 1.2. to 1.7), while the compact UVLGs have the
blue colors ($FUV-r \sim$ 0.5 to 2) and small 4000 \AA\ break
strengths (D4000 $\sim$ 1.0 to 1.3) of very young stellar populations
(e.g. Kauffmann et al 2003a). The UV-optical colors of the compact
UVLGs are very similar to those of LBGs (S01,P01,G02).

\subsection{Chemical and Kinematic Properties}

The metallicities in the ionized gas in the UVLGs are typical for SDSS
galaxies of the corresponding stellar mass (Tremonti et al. 2004): the
large (massive) UVLGs have oxygen abundances of several times solar,
while the compact (less massive) UVLGs span a broad range in
metallicity from $\sim$30\% solar to several times solar. We note that
the ``strong-line'' methods like those used by Tremonti et al. to
estimate metallicity are known to yield systematically higher
metallicities than direct techniques (Kennicutt, Bresolin,
\& Garnett 2003). Pettini \& Pagel (2004) have recently recalibrated
the ``[OIII]/[NII]'' strong line metallicity indicator. Using their
calibration, the UVLG metallicities would be roughly solar for the
large UVLGs and $\sim$30\% solar to solar for the compact UVLGs. These
latter metallicities are similar to estimates for LBGs (Pettini et
al. 2001; Shapley et al. 2004).

Typical emission-line velocity
dispersions range from 50 to 150 km sec$^{-1}$, with little dependence
on galaxy mass. These are similar to
values in LBGs (Pettini et al. 2001). Using these velocity
dispersions and the u-band half-light radii we have computed dynamical
masses ($M_{dyn} = 5 \sigma_{gas}^2 R_u/G$). These agree with the
stellar masses to within a typical factor of three.

\subsection{Presence of Type 2 AGN}

We have used the emission-line flux ratios [OIII]/H$\beta$ vs.
[NII]/H$\alpha$ to identify UVLGs in which a type 2 AGN is making a
significant contribution to the emission-line spectrum (see Kauffmann
et al. 2003b).
\footnote{Kauffmann et al. (2003b) show that the SDSS spectra and
images are dominated by starlight in type 2 AGN, so our methods for
inferring galaxy masses, star formation rates, sizes, etc. are not
significantly affected by the AGN.}
We find that AGN are present in 35\% of the large
UVLGs, and in 15\% of the compact UVLGs (the stronger emission lines
produced by the more intense star formation makes the presence of an AGN
more difficult to detect in the compact UVLGs).
The great majority
(83\%) of the AGN would be classified by Kauffmann et al. as
``Composite'' systems in which the observed emission-lines are
contributed by a combination of an AGN and O stars. The high rate of
AGN in UVLGs is not surprising given the link between strong AGN and
star formation in the population of massive galaxies in the SDSS
(Kauffmann et al. 2003b).

\section{Discussion \& Conclusions}

The large UVLGs have specific star formation rates sufficient to build
their stellar mass over a Hubble time. Thus, they are not starburst
systems. Instead they owe their high UV luminosities (high star
formation rates) to their relatively large masses. These systems
appear to be the tail of the distribution of the normal SDSS high-mass
galaxy population, extending to high star formation rates, low stellar
surface mass densities, large radii, low concentrations, and 
(by implication) high
gas-mass fractions. That is, these appear to be among the most massive
late type galaxies in the current epoch.  They are rare objects:
comparably massive galaxies today are primarily old, gas-poor,
bulge-dominated systems (Kauffmann et al. 2003c).

In contrast, the compact UVLGs overlap the properties of the Lyman
Break Galaxies in all respects considered in this paper (see Table
1). 
By definition, they have similar UV luminosities.  They have
sizes, surface brightnesses, UV extinctions, and implied star
formation rates that all overlap those of the LBGs.  They have similar
stellar masses, velocity dispersions, and gas-phase metallcities as
LBGs. Their rest-frame UV-optical spectral energy distributions are
similar to those of LBGs. Finally, they are forming stars at a rate
sufficient to form their present stellar mass in typically 1 to 2
Gigayear. The compact UVLGs are clearly systems
undergoing starbursts, and a relatively low duty cycle for
such bursts presumably contributes to their relative rarity today.
Since the compact UVLGs do appear similar to the LBGs, these ``living
fossils'' may provide an opportunity for a detailed {\it local}
investigation of the same physical processes that occurred in typical
star forming galaxies in the early universe.

\acknowledgments
GALEX (Galaxy Evolution Explorer) is a NASA Small Explorer, launched in April 2003.
We gratefully acknowledge NASA's support for construction, operation,
and science analysis for the GALEX mission,
developed in cooperation with the Centre National d'Etudes Spatiales
of France and the Korean Ministry of 
Science and Technology. Funding for the creation and distribution of
the SDSS Archive has been provided by the Alfred P. Sloan Foundation,
the Participating Institutions, NASA, NSF, DoE, Monbukakusho, and
the Max Planck Society.




\clearpage



\begin{figure}
\epsscale{1.0}
\plotone{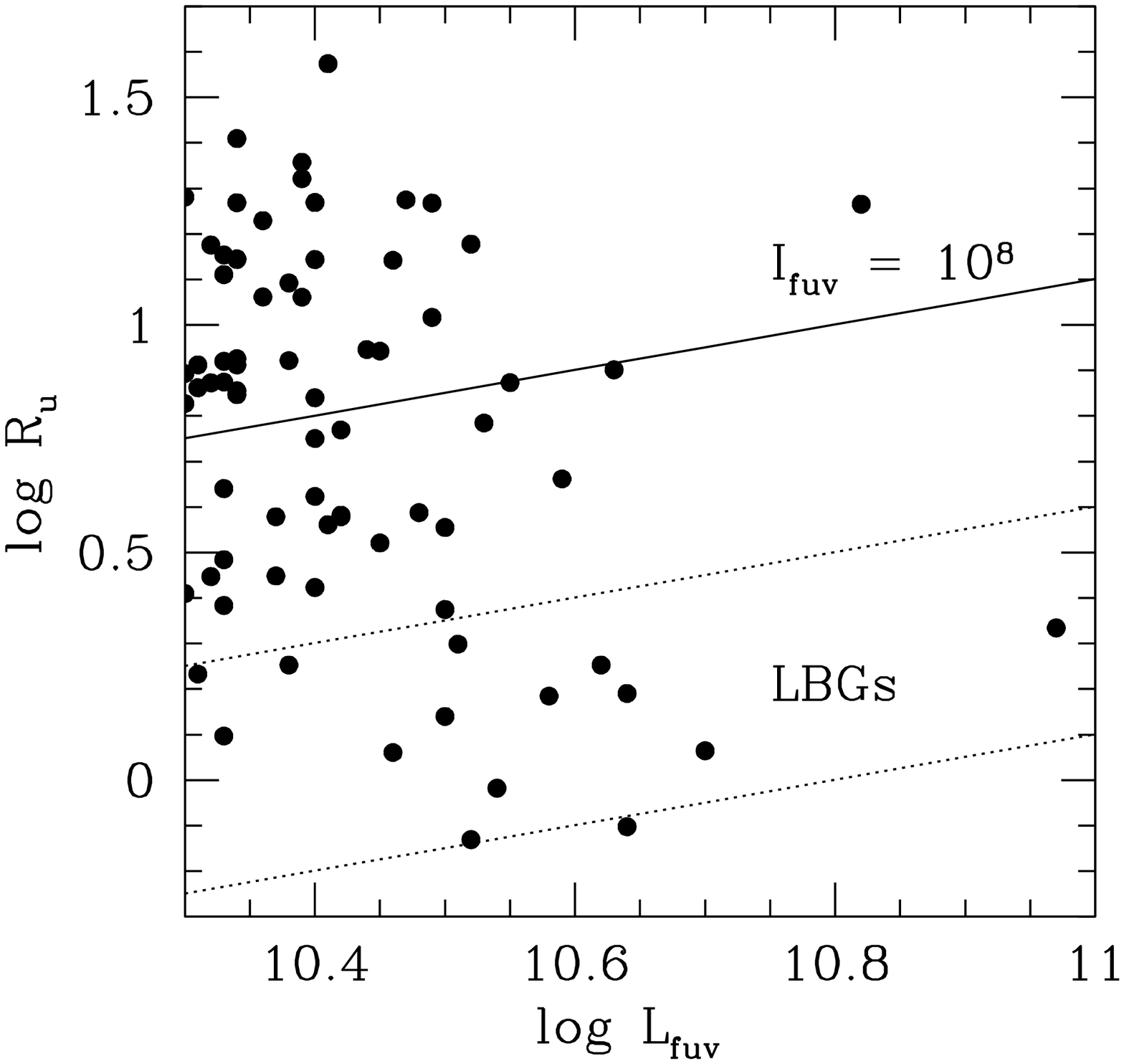}
\caption{Far-Ultraviolet luminosity ($L_{\odot}$)
{\it vs.} the seeing-deconvolved galaxy half-light radius
measured in the SDSS u-band (kpc). We define
large (compact) UVLGs as those galaxies
with far-UV surface brightnesses less than (greater than) 
10$^8 L_{\odot}$ kpc$^{-2}$ (shown by the solid line). The location
of typical high-z Lyman Break Galaxies is bounded by the dotted lines.
\label{fig1}} 
\end{figure}

\begin{figure}
\epsscale{1.0}
\plotone{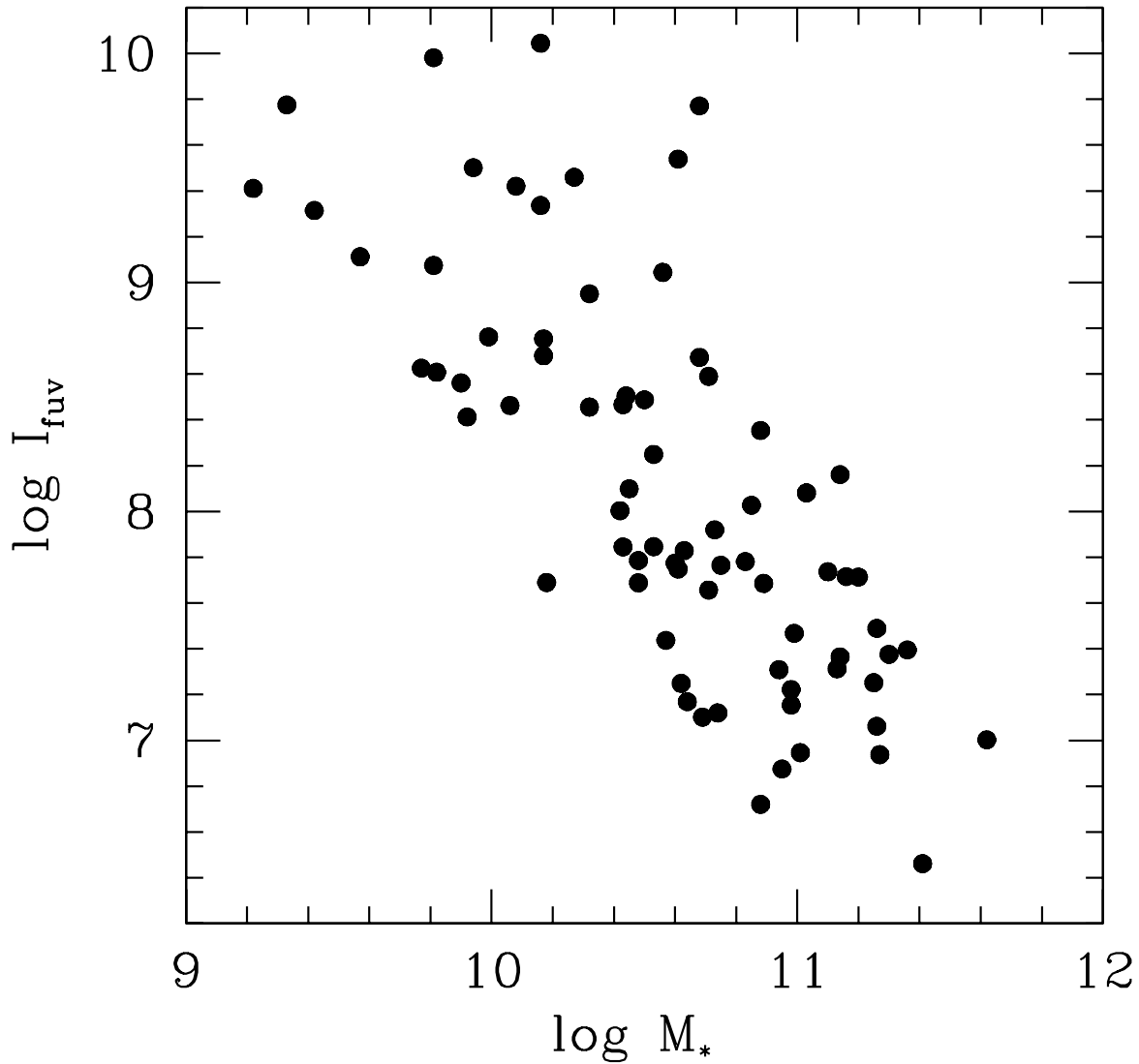}
\caption{Far-Ultraviolet effective surface brightness (the mean surface
brightness interior to the SDSS u-band half light radius) in units of
$L_{\odot}$ kpc$^{-2}$
{\it vs.} the stellar
mass ($M_{\odot}$) for the UVLGs. 
We define large (compact) UVLGs as those galaxies
with surface brightnesses less than (greater than) 10$^8 L_{\odot}$ kpc$^{-2}$.
The strong correlation between mass and surface brightness means that
compact UVLGs are typically much less massive than large UVLGs
($\sim10^{10}$ {\it vs.} $\sim10^{11} M_{\odot}$ respectively).
Typical Lyman Break Galaxies lie in the upper left of this plot
($log~I_{fuv} \sim$
9 to 10 and $log~M_* \sim$ 9 to 11).
\label{fig2}} 
\end{figure}

\begin{figure}
\epsscale{1.0}
\plotone{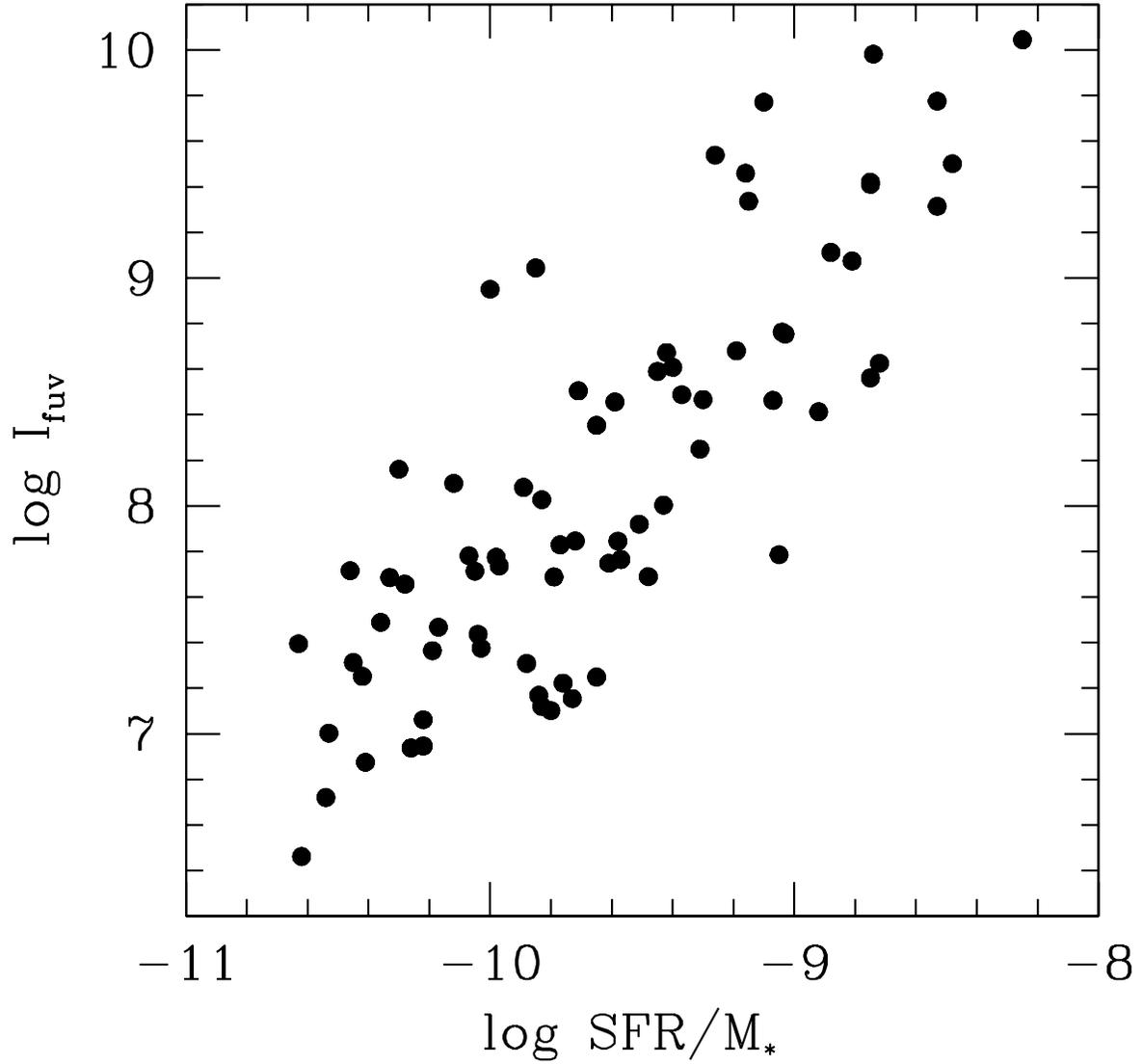}
\caption{Far-Ultraviolet effective surface brightness {\it vs.} the star
formation rate per unit mass (year$^{-1}$). The 
compact UVLGs are typically starbursts ($M_*/SFR$ is of-order a Gigayear),
while the large UVLGs are not ($M_*/SFR$ is of-order the Hubble time).
Typical Lyman Break Galaxies lie in the upper right of this plot
($log~I_{fuv} \sim$ 9 to 10 and $log~SFR/M_* \sim$ -9 to -8).
\label{fig3}}
\end{figure}

\clearpage

\begin{deluxetable}{lcccc}
\tabletypesize{\scriptsize}
\tablecaption{Summary of Properties}
\tablewidth{0pt}
\tablehead{
\colhead{Parameter} & \colhead{Units} &\colhead{Large UVLGs}
& \colhead{Compact UVLGs} & 
\colhead{Lyman Break Galaxies}\\
\colhead{(1)}&\colhead{(2)}&
\colhead{(3)}&\colhead{(4)}&\colhead{(5)}
}
\startdata
$log~L_{fuv}$ & $L_{\odot}$ & 10.3 -- 10.5 & 10.35 -- 10.65 & 10.3 -- 11.3\\
$log~R50_u$   & kpc & 0.9 -- 1.3 & 0.0 -- 0.7 & 0.0 -- 0.5\\
$log~I_{fuv}$ & $L_{\odot}$/kpc$^2$ & 6.9 -- 7.8 & 8.2 -- 9.8 & 9 -- 10\\
$log~M_*$     & $M_{\odot}$ & 10.5 -- 11.3 & 9.5 -- 10.7 & 9.5 -- 11\\
$log~M_{dyn}$ & $M_{\odot}$ & 10.4 -- 11.6 & 10.0 -- 10.8 & 10.0 -- 10.5\\
$log~\mu_*$   & $M_{\odot}$/kpc$^2$ & 7.9 -- 8.7 & 8.0 -- 9.1 & 8.5 -- 9.0\\
$A_{fuv}$     & mag & 0.3 -- 2.0 & 0.6 -- 2.1 & 1 -- 3\\
$log~SFR$         & $M_{\odot}$/year & 0.6 -- 1.2 & 0.6 -- 1.4 & 0.5 -- 2.5\\
$log~SFR/M_*$     & year$^{-1}$ & -10.5 -- -9.5 & -9.8 -- -8.6 & -9 -- -8\\
$(FUV-r)$     & AB mag & 1.8 -- 2.9 & 0.6 -- 2.1 & 0.2 -- 2.2\\
$D(4000)$     & -- & 1.2 -- 1.7 & 1.0 -- 1.3 & N/A\\
$12+log~O/H$  & -- & 8.55 - 8.75 & 8.2 - 8.7 & 7.7 - 8.8\\
$log~\sigma_{gas}$ & km/sec & 1.7 - 2.1 & 1.8 - 2.2 & 1.7 - 2.1\\
\enddata

\tablecomments{For the UVLGs, the entries in each column refer to
the 10 to 90 percentile range. 
For the Lyman Break Galaxies, the
ranges are just meant to be representative
of the population at $z \sim$ 3. These
were extracted from Shapley et al. (2001), 
Papovich et al. (2001), Pettini et al (2001), Giavalisco (2002), 
Ferguson et al (2004), and Giavalisco (private communication).
All the parameters are described in the text.}
\end{deluxetable}

\end{document}